\documentclass[twocolumn,showpacs,preprintnumbers,amsmath,amssymb,aps]{revtex4}
\usepackage{graphicx}
\usepackage{bm}
\usepackage{amstext}

\begin{document}
\preprint{}

\title{$\Lambda-$hyperon interaction with nucleons}
\author{M. Ikram$^1$} 
\author{S. K. Singh$^2$} 
\author{S. K. Biswal$^2$} 
\author{M. Bhuyan$^2$} 
\author{S. K. Patra$^2$} 
\affiliation{$^1$Department of Physics, Aligarh Muslim University, Aligarh -
202002, India} 
\affiliation{$^2$Institute of Physics, Sachivalaya Marg, Bhubaneswar -
751005, India.} 

\date{\today}

\begin{abstract}
We study the interaction of $\Lambda-$hyperon with proton and neutron inside 
a nucleus within the framework of relativistic mean field formalism. 
The single particle energy levels for some of the specific proton and neutron 
orbits are analyzed with the addition of $\Lambda-$successively. 
We found that the interaction of $\Lambda$ with neutron is more stronger 
than proton.
\end{abstract}


\maketitle


\vspace{0.8cm}

Hypernucleus provides an opportunity to enter into the strangeness world.
After introducing the strangeness degree of freedom to bound nuclear 
system, the multi-baryonic system avails $\Lambda$-nucleon ($\Lambda N$) 
interaction in addition to nucleon-nucleon (NN) interaction.
However, $\Lambda N$ interaction is weaker than NN 
but it is imperative as well as important to describe the strange system.
Many of the theoretical calculations have been made to 
give the importance of $\Lambda N$ interaction and 
to facilitate the path toward multi-strange systems
~\cite{bodmer1962,gibson1988,gibson1994,schaffner1994,gibson2013}.

The information about the hyperon-nucleon interaction especially 
$\Lambda p$ and $\Lambda n$ can be extracted from the hypernuclei. 
Due to zero isospin of $\Lambda-$hyperon, the one pion exchange is 
prohibited in $\Lambda N$ interaction. 
Which means the $\Lambda N$ interaction is governed by two pion exchange.
In this way, $\Lambda N-\Sigma N$ coupling plays a significant role 
to drive the $\Lambda N$ interaction.
The most interested mirror hypernuclei are $^4_\Lambda$H and $^4_\Lambda$He which 
reflect the difference in strength of $\Lambda p$ and $\Lambda n$ interactions.
The $\Lambda N$ interaction occurs via $\Lambda N-\Sigma N$ coupling where,  
$\Lambda p$ couples to $\Sigma^+n$ and $\Lambda n$ is coupled with 
$\Sigma^-p$~\cite{gibson1995}.
Therefore, $\Sigma N$ plays a role as an intermediate state to access the 
$\Lambda N$ interaction.
On the basis of this coupling, it is expected that the mass difference 
of $\Sigma^+-\Sigma^-$ is responsible to make a difference in the 
strength of $\Lambda p$ and $\Lambda n$ interactions
as discussed in Refs.~\cite{gibson1988,gibson1995,dalitz1964}.
The difference in strength of these interactions is a direct consequence of 
charge symmetry breaking which is observed in mirror 
hypernuclei~\cite{gibson1988,gibson1995,dalitz1964}.
It is well understood that any kinds of change take place in 
the interaction is directly reflected in nuclear potential as well as 
single-particle energy.
Thus, it is very much interesting to analyze the single-particle 
energy or potential to study the net effect on interaction, either 
with the addition of $\Lambda-$hyperon or any other effects.
In this work, we study the single-particle energy as well as 
potentials of some medium and superheavy hypernuclei to 
demonstrate this mechanism by employing 
the relativistic mean field (RMF) formalism.

Recently, the RMF theory is quite successful for studying 
the finite and infinite nuclear systems.
Quite successful to study the equation of state (EOS) for normal as 
well as high dense neutron matter.
Since neutron star is a compact object with nuclear density 
$\rho=(8-10)\rho_0$, where $\rho_0$ is the nuclear matter density at saturation, 
so there must be the possibility of formation of strange baryon.
In this context, addition of strangeness degree of freedom to RMF formalism is 
obvious for the suitable expansion of the model and this type of attempts have 
already been made~\cite{schaffner1994,rufa1990,glendenning1993,schaffner1993,
mares1994,sugahara1994,vretenar1998,schaffner2002,lu2003,shen2006,win2008}.

The relativistic mean field Lagrangian density for single-$\Lambda$ 
hypernuclei has been given in Refs.~\cite{rufa1990,glendenning1993,mares1994,
sugahara1994,vretenar1998,lu2003,win2008}. 
To study the multi-strange system in quantitative way, the additional 
strange scalar ($\sigma^*$) and vector ($\phi$) mesons have been included 
which simulate the $\Lambda \Lambda$ interaction
~\cite{schaffner1994,schaffner1993,schaffner2002,shen2006}. 
Now, the total Lagrangian density can be written as
\begin{eqnarray} 
\mathcal{L}&=&\mathcal{L}_N+\mathcal{L}_\Lambda+\mathcal{L}_{\Lambda\Lambda} \;, 
\end{eqnarray}
\begin{eqnarray}
{\cal L}_N&=&\bar{\psi_{i}}\{i\gamma^{\mu}
\partial_{\mu}-M\}\psi_{i}
+{\frac12}(\partial^{\mu}\sigma\partial_{\mu}\sigma
-m_{\sigma}^{2}\sigma^{2})		   
-{\frac13}g_{2}\sigma^{3}                  \nonumber \\
&&-{\frac14}g_{3}\sigma^{4}
-g_{s}\bar{\psi_{i}}\psi_{i}\sigma 		
-{\frac14}\Omega^{\mu\nu}\Omega_{\mu\nu}
+{\frac12}m_{\omega}^{2}\omega^{\mu}\omega_{\mu}		\nonumber \\
&&-g_{\omega }\bar\psi_{i}\gamma^{\mu}\psi_{i}\omega_{\mu}    
-{\frac14}B^{\mu\nu}B_{\mu\nu} 
+{\frac12}m_{\rho}^{2}{\vec{\rho}^{\mu}}{\vec{\rho}_{\mu}}  
-{\frac14}F^{\mu\nu}F_{\mu\nu}                        \nonumber \\
&&-g_{\rho}\bar\psi_{i}\gamma^{\mu}\vec{\tau}\psi_{i}\vec{\rho^{\mu}} 
-e\bar\psi_{i}\gamma^{\mu}\frac{\left(1-\tau_{3i}\right)}{2}\psi_{i}A_{\mu}\;,  \nonumber \\
\mathcal{L}_{\Lambda}&=&\bar\psi_\Lambda\{i\gamma^\mu\partial_\mu
-m_\Lambda\}\psi_\Lambda
-g_{\sigma\Lambda}\bar\psi_\Lambda\psi_\Lambda\sigma  
-g_{\omega\Lambda}\bar\psi_\Lambda\gamma^{\mu}\psi_\Lambda \omega_\mu \;,  \nonumber \\
\mathcal{L}_{\Lambda\Lambda}&=&{\frac12}(\partial^{\mu}\sigma^*\partial_{\mu}\sigma^*  
-m_{\sigma^*}^{2}\sigma^{*{2}})
-{\frac14}S^{\mu\nu}S_{\mu\nu}
+{\frac12}m_{\phi}^{2}\phi^{\mu}\phi_{\mu}    \nonumber \\
&&-g_{\sigma^*\Lambda}\bar\psi_\Lambda\psi_\Lambda\sigma^*
-g_{\phi\Lambda}\bar\psi_\Lambda\gamma^\mu\psi_\Lambda\phi_\mu \;,      
\end{eqnarray}
where $\psi$ and $\psi_\Lambda$ denote the Dirac spinors for 
nucleon and $\Lambda-$hyperon, whose masses are M and
$m_\Lambda$ respectively.
Because of zero isospin, the $\Lambda-$hyperon does not couple 
to ${\rho}$- mesons.
The quantities $m_{\sigma}$, $m_{\omega}$, $m_{\rho}$, $m_{\sigma^*}$, 
$m_{\phi}$ are the masses of included mesons and $g_s$, $g_{\omega}$, 
$g_{\rho}$, $g_{\sigma\Lambda}$, $g_{\omega\Lambda}$, $g_{\sigma^*\Lambda}$, 
$g_{\phi\Lambda}$ are their coupling constants. 
The nonlinear self-interaction coupling of ${\sigma}$ mesons is denoted 
by $g_2$ and $g_3$. 
The total energy of the system is given by 
$E_{total} = E_{part}(N,\Lambda)+E_{\sigma}+E_{\omega}+E_{\rho}
+E_{\sigma^*}+E_{\phi}+E_{c}+E_{pair}+E_{c.m.},$
where $E_{part}(N,\Lambda)$ is the sum of the single particle energies of the 
nucleons (N) and hyperon ($\Lambda$).
The energies parts $E_{\sigma}$, $E_{\omega}$, $E_{\rho}$, $E_{\sigma^*}$, 
$E_{\phi}$, $E_{c}$, $E_{pair}$ and $E_{cm}$ are the contributions of meson 
fields, 
Coulomb field, pairing energy and the center-of-mass energy, respectively.
For present study, we use the NL3* parameter set through out the 
calculations~\cite{lalazissis09}. 
To find the numerical values of used $\Lambda-$meson coupling constants, 
we adopt the relative coupling for $\sigma$, $\omega$, $\sigma^*$ and $\phi$ fields.
The ratio of meson-hyperon coupling to meson-nucleon coupling is defined as 
$R_\sigma=g_{\sigma\Lambda}/g_s$ and $R_\omega=g_{\omega\Lambda}/g_\omega$
$R_{\sigma^*}=g_{\sigma^*\Lambda}/g_s$ and $R_\phi=g_{\phi\Lambda}/g_\omega$.
The relative coupling values are used as $R_\omega=2/3$, 
$R_\phi=-\sqrt{2}/3$, $R_\sigma=0.621$
and $R_{\sigma^*}=0.69$~\cite{dover1984,schaffner1994,mares1994,keil2000}.
In present calculations, we use the constant gap BCS approximation to 
include the pairing interaction and the centre of mass 
correction is included by  $E_{cm}=-(3/4)41A^{-1/3}$.
 
\begin{table*}
\caption{\label{tab1}The calculated total and per particle binding energy for 
single-$\Lambda$ hypernuclei and their normal counter parts are listed here. 
The single-$\Lambda$ binding energy for s- and p-state of considered hypernuclei 
are also mentioned and compared with available exp. values which are given in 
parentheses~\cite{hashimoto2006}. 
The radii are also displayed. Energies are given in MeV and radii are in fm.}
\renewcommand{\tabcolsep}{0.39cm}
\renewcommand{\arraystretch}{1.2}
\begin{tabular}{lccccccc}
\hline
\hline
&BE &$B_\Lambda^s$&$B_\Lambda^p$ &$r_{ch}$ &$r_{rms}$   &$r_n$ &$r_\Lambda$ \\
\hline
$^{48}$Ca            &414.2  &                    &                   &3.444 &3.496&         3.591&              \\
$^{48}_\Lambda$Ca    &428.2  &21.7                &13.4               &3.435 &3.454&         3.554&         2.708\\
$^{208}$Pb           &1639.3 &                    &                   &5.499 &5.624&         5.736&              \\
$^{208}_\Lambda$Pb   &1660.1 &26.9(26.3$\pm$0.8)  &23.0(21.3$\pm$0.7) &5.490 &5.602&         5.718&         4.011\\
$^{298}$114          &2119.3 &                    &                   &6.248 &6.397&         6.513&              \\
$^{298}_\Lambda$114  &2168.6 &27.2                &24.1               &6.241 &6.378&         6.497&         3.209\\
\hline
\hline
\end{tabular}
\end{table*}

The addition of $\Lambda-$hyperon to normal nuclei enhances the 
binding and shrinks the core of the system.
This happens because of glue like behaviour of $\Lambda-$hyperon.
These observations are shown in Table 1, where the total binding 
energy (BE) of hypernuclei are larger than their normal counter parts 
and a reduction in total radius ($r_{rms}$) of hypernuclei is also observed. 
For example, the total radius of $^{48}$Ca is 3.496 fm, which is reduced 
to 3.454 fm by addition of single $\Lambda$ into $^{48}$Ca nucleus.
The increasing value of single-$\Lambda$ binding energy ($B_\Lambda$) for 
$s-$state from medium to superheavy hypernuclei confirming the potential 
depth of lambda particle in nuclear matter which would be -28~MeV
~\cite{schaffner1994,millener1988}.

The lambda hyperons are introduced into the nucleus to see the effects 
of substituted hyperons on single particle energies of neutron as well as proton.
We choose the system in such a way to cover the range from medium to superheavy 
hypernuclei, for example, $^{48}_{n\Lambda}$Ca, $^{208}_{n\Lambda}$Pb and 
$^{298}_{n\Lambda}$114.
To study the single particle energy, we choose some specific energy levels 
(first occupied and other higher orbital) for both neutron as well as proton.
Initially, in normal nuclei the higher orbitals of proton and neutron are 
fully occupied but due to successive addition of $\Lambda-$hyperons with 
replacing neutrons the upper neutron orbitals become unfilled however 
proton orbitals are still occupied.
In this way, we analyze the energy space for first filled and other higher 
levels with successive addition of $\Lambda-$hyperons.
In $^{48}_{n\Lambda}$Ca, we study the behaviour of proton (1$s_{1/2}, 
1d_{3/2}$) and neutron ($1s_{1/2}, 1f_{7/2}$) energy levels in the respect 
of substituted $\Lambda$'s.
It is evident from Fig. 1(a, b, c) that the neutron levels go dipper with 
increasing number of substituted $\Lambda-$hyperons. 
The behaviour of first occupied proton level looks to be in same trend as 
neutron but feels small attraction comparable to neutron.
This observation reflects that the hyperon interact more strongly 
with neutron in comparison to proton.
However, the nature of both the interactions is attractive and the 
present outcome confirms that $\Lambda n$ interaction is more 
stronger than  $\Lambda p$.
Again it is found that, the first occupied orbital of neutron is 
more effective by addition of $\Lambda$'s compared to higher orbitals. 
For example, $1s_{1/2}(n)$ orbital feels more attraction in comparison 
to $1f_{7/2}(n)$ levels in $^{48}_{n\Lambda}$Ca hypernuclei.
This may be expected because of $\Lambda-$hyperon resides at the center 
of the  nucleus for most of the time and attracts the surrounding 
nucleons towards the centre.
On the other hand, the higher proton and neutron levels 
are in opposite trend to each other.
In $^{48}_{n\Lambda}$Ca hypernuclei, the neutron level ($1f_{7/2}(n)$) 
feels a small attraction with the addition of successive $\Lambda$'s 
while the proton level ($1d_{3/2}(p)$) seems in opposite trend. 
This may be because of the Coulomb repulsion.
The injected $\Lambda-$hyperons reduce the isospin of the whole system 
and as a result the Coulomb interaction becomes more effective.
This mechanism can be explained by higher proton orbital, where it goes toward 
less bound nature, for example, $1d_{3/2}(p)$ level in $^{48}_{n\Lambda}$Ca, 
$1h_{11/2}(p)$ level in $^{208}_{n\Lambda}$Pb and $1i_{13/2}(p)$ level 
in $^{298}_{n\Lambda}$114 as shown in Fig. 1(a, b, c).
It is also to be noted that the system with strangeness feels to be more 
bound  with the addition of $\Lambda-$hyperon succesively, and 
single particle energy levels goes dipper and dipper up to a 
certain number of hyperons.
Beyond this limit the nature of hypernuclear system becomes 
reversed and would be collapsed as reffered as key point 
in Ref.~\cite{schaffner1993}.
This behaviour can be noticed in $^{48}_{n\Lambda}$Ca, where after the 
addition of 14 $\Lambda$'s, the neutron and proton potentials are reduced.
The same behavior of single particle energy levels is observed not 
only for $^{208}_{n\Lambda}$Pb but $^{298}_{n\Lambda}$114 superheavy 
hypernucleus also as shown in Fig. 1(a, b, c).
 
\begin{figure*}
\vspace{1.2cm}
\includegraphics[scale=0.69]{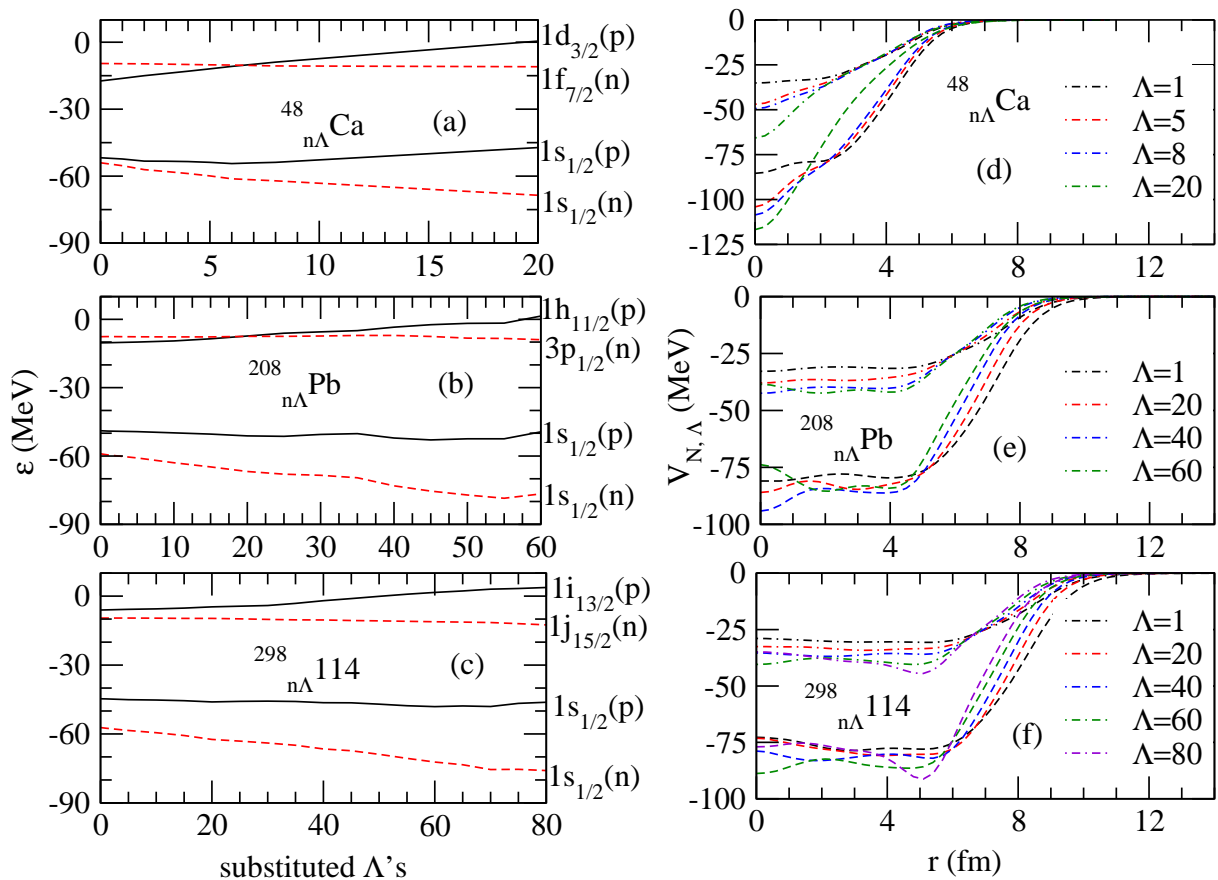}
\caption{\label{fig1}(color online) The first occupied and higher orbits of 
neutron and proton are shown for $^{48}_{n\Lambda}$Ca, $^{208}_{n\Lambda}$Pb 
and $^{298}_{n\Lambda}$114 hypernuclei as a function of substituted $\Lambda$'s 
with replacing neutrons on left portion of figure where, 
the dashed lines in red color represent the neutron levels while the
proton levels are represented by solid black lines.
The neutron ($V_N=V_\sigma+V_\omega+V_\rho$) and 
lambda ($V_\Lambda=V_{\sigma\Lambda}+V_{\omega\Lambda}$) mean potentials 
are shown for considered hypernuclei in right portion of the figure. 
The dashed and dash dotted lines with different colors represent the 
neutron and lambda mean potential, respectively. 
}
\end{figure*}

The neutron ($V_N$) and lambda ($V_\Lambda$) mean potentials are plotted 
for the considered hypernuclei in Fig. 1(d, e, f).
From this figure, it is reveal that how does the potential shape and depth 
are affcted by successive addition of $\Lambda-$hyperons to the core 
of the nuclei.
The behaviour of the potential is plotted only for some number of hyperons.
For example, the neutron and lambda mean potentials are given for 
$\Lambda=$ 1, 5, 8 and 20 for $^{48}_{n\Lambda}$Ca. 
In the same way for superheavy hypernuclei, $^{298}_{n\Lambda}$114, the 
potentials are shown for $\Lambda=$ 1, 20, 40, 60 and 80.
It is shown in Fig. 1d, that the $V_N$ gets the depth around -85~MeV with 
addition of one $\Lambda$ to $^{48}$Ca nucleus and it goes dipper with the 
addition of more $\Lambda$'s.
Look into the $^{208}_{n\Lambda}$Pb, both the potential depth 
($V_N$ and $V_\Lambda$) goes dipper with inclusion of 40 $\Lambda$'s but 
revert on addition of 60 $\Lambda$'s.
The reason is simple because there is some limitation of 
numbers of hyperons for a particular system to make the maximum binding.
The potentials ($V_N$ and $V_\Lambda$) are distorted with the 
addition of 20 $\Lambda$'s to the core of $^{48}$Ca nucleus. 
It is clearly seen that the depth and shape of lambda mean potential 
is completely affected with increasing the $\Lambda-$interactions with nucleons.

The present study is fully devoted to demonstrate the difference in 
strength of $\Lambda$-nucleon interactions in multi-baryonic system. 
The results show that the $\Lambda n$ interaction is more stronger than 
$\Lambda p$ which is in agreement with scattering data.
We also observe the affects on $V_N$ and $V_\Lambda$ mean potentials 
by introducing the $\Lambda-$hyperons successively.
In our study we find that the binding of nuclear system is increasing 
with increasing the injected hyperons but up to a certain numbers. 
The injection of all other hyperons in presence of $\Lambda$ 
would provide an opportunity to study the high dense system  
where cluster of baryon octet or pure hyperonic matter may exist.
This type of study will be helpful to simulate the structure of 
astrophysical objects like, neutron or hyperon stars.
Work is in progress in this direction.

One of the author (MI) wishes to acknowledge the hospitality provided by 
Institute of Physics, Bhubaneswar during the work.


\end{document}